# Transmodal Fabry-Pérot Resonance: Theory and Realization with Elastic Metamaterials

Joshua Minwoo Kweun,[1] Hyung Jin Lee,[2] Joo Hwan Oh,[3] Hong Min Seung,[4]
and Yoon Young Kim[1,2,*]

[1] *School of Mechanical and Aerospace Engineering, Seoul National University, 1 Gwanak-ro, Gwanak-gu, Seoul 08826, South Korea*
[2] *Institute of Advanced Machines and Design, Seoul National University, 1 Gwanak-ro, Gwanak-gu, Seoul 08826, South Korea*
[3] *School of Mechanical and Nuclear Engineering, Ulsan National Institute of Science and Technology, UNIST-gil 50, Eonyang-eup, Ulju-gun, Ulsan 44919, South Korea*
[4] *Center for Safety Measurement, Division of Metrology for Quality of Life, Korea Research Institute of Standards and Science (KRISS), 267 Gajeong-ro, Yuseong-gu, Daejeon 34113, South Korea*

We discovered a new transmodal Fabry-Pérot resonance that one elastic-wave mode is maximally transmitted to another when the phase difference of two dissimilar modes through an anisotropic layer is exactly odd multiples of $\pi$. Unlike the well-established Fabry-Pérot resonance, the transmodal resonance must involve two coupled elastic-wave modes, longitudinal and shear. The formation of wiggly transmodal transmission spectra is due to structural instability appearing in anisotropic mode-coupled elastic-media. Experiments with elastic metamaterials confirmed our findings which can play a critical role in shear-mode ultrasound applications.

The Fabry-Pérot resonance (FPR) is a well-known wave phenomenon taking place in a single or multiple layers of different media [1-3]. For transmission over a single layer, an incident wave can be completely transmitted at FPRs due to the constructive interference of multiply-reflected waves inside the layer with the thickness equal to multiples of the half wavelength. We will call the conventional FPR the unimodal FPR (UFPR) because it concerns a single wave mode, either longitudinal or shear modes alone. Unlike in electro-magnetic (transverse modes only) and acoustic (a longitudinal mode only) waves, both longitudinal and transverse modes, which are called longitudinal (L) and shear (S) modes, respectively, exist for elastic waves [4,5] due to the atomic bindings in a solid state. However, the UFPR condition for elastic waves is only possible if a layer is isotropic or anisotropic with its principal axis aligned with the wave direction. In general anisotropic elastic layers, longitudinal and transverse particle-motions can be coupled together during wave propagation, and thus wave transmission through this mode-coupled layer causes power exchange (or conversion) between the L and S modes with wiggly frequency responses. Some studies have been carried out on the intriguing mode-coupled wave behavior in elastic media [6-9]. However, a phenomenon that one wave mode is maximally transmitted to another mode, which we call the transmodal Fabry-Pérot resonance (TFPR) phenomenon, has not been explored yet.

Because conventional transducers are not very effective for shear wave transduction, there is a strong demand for efficient generation and measurement of them. In this respect, the mode conversions from L-to-S or S-to-L waves possibly by the TFPRs can open an avenue for shear wave handling because the transduction of the L mode is rather efficient. In particular, the L-to-S (mode) conversion has a great potential in medical and industrial ultrasonic applications [10-14]. For example, the S mode can provide higher resolution for the ultrasonic non-destructive testing and higher ultrasound transmission for trans-skull measurements and treatments than the L mode [10,12,13,15]. Unfortunately, the selective excitation of the S mode using conventional transducers has been seriously limited due to the intrinsic problem of piezoelectric materials. Despite the high demands for S mode waves, the L-to-S conversion has been realized only by using Snell's critical angle, resulting in low transmission power and high dependence on material properties [4,6,7,10]. Therefore, an alternative efficient L-to-S conversion method is needed.

In this Letter, we rigorously explored the TFPR and established the condition for the TFPR at which an incident L mode can be maximally converted to an S mode, and vice versa, by an anisotropic mode-coupled layer. This idea is sketched in Fig. 1(a). The illustrated mode-coupled layer was actually realized by elastic metamaterials (EMMs). EMMs are composite elastic materials with artificial microstructures made to exhibit unusual wave characteristics such as cloaking and negative refraction [6,7,16-18], as a counterpart of electro-magnetic metamaterials [19,20]. Extreme anisotropy as well as a wide range of material property has been realized by EMMs [16-18,21]. The specific unit cell used to make our metamaterial will be given later with experiments.

For our analysis, we consider plane harmonic elastic waves in a one-dimensional regime in which the phase velocity and wave mode polarization can be determined by the following Christoffel equation [4]:

$$k^2 \mathbf{\Gamma} \mathbf{v} = k^2 \begin{bmatrix} C_{11} & C_{16} \\ C_{16} & C_{66} \end{bmatrix} \begin{bmatrix} v_x \\ v_y \end{bmatrix} = \rho \omega^2 \begin{bmatrix} v_x \\ v_y \end{bmatrix}, \tag{1}$$

where $\mathbf{\Gamma}$ is the Christoffel matrix for unidimensional elastic waves consisting of stiffness tensors $C_{ij}$, and $k$ is the wave number, and $\mathbf{v}$ is the polarization velocity vector, and $\rho$ is the mass density of the mode-coupled medium. Among $C_{ij}$'s, $C_{11}$ and $C_{66}$ correspond to the longitudinal and shear stiffness, respectively, while $C_{16}$, to mode-coupling stiffness. Clearly, mode-coupling can occur as long as $C_{16} \neq 0$.

Figure 1(b) shows transmission spectra through mode-coupled media ($C_{16} \neq 0$) for various values of $C_{11}/C_{66}$. The L-to-S transmodal transmission or mode conversion rate becomes maximized at certain frequencies, which we call the TFPR frequencies. The figure also shows that the maximum L-to-L (mode) transmission occurs at the anti-TFPR which does not always satisfy the UFPR condition ($kd = m\pi$, $m$: integer). The relative ratio of the L-to-L transmission to the L-to-S conversion under L-mode incidence varies considerably depending on the value of $C_{11}/C_{66}$ as shown in Fig. 1(b). The L-to-L transmission can even vanish when $C_{11} = C_{66}$. Figure 1 (b) shows peculiar wiggly transmission spectra which cannot be observed in the unimodal FPR [1,3,22]. We will show that it is related to the so-called structural instability of mode-coupled media.

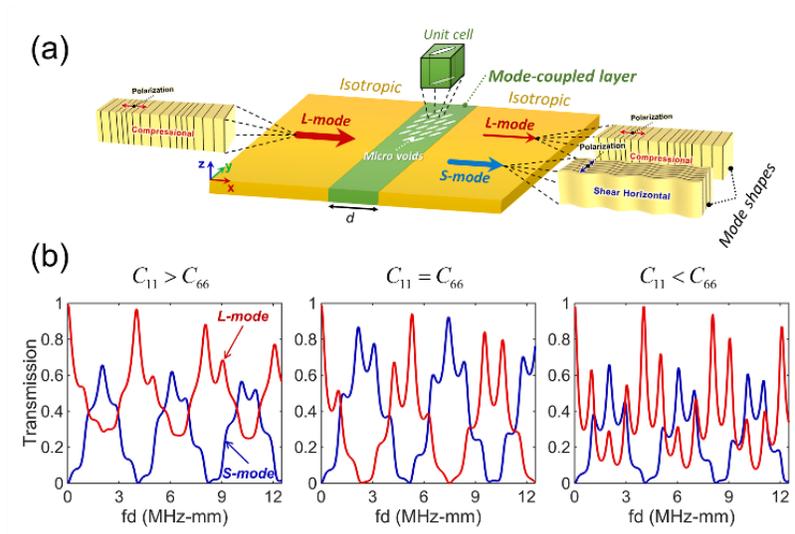

FIG. 1 (color online). (a) A schematic illustration of the TFPR in a mode-coupled layer surrounded by an isotropic medium. At TFPRs, an incident L mode can be maximally converted to the S mode as it propagates through a certain class of anisotropic elastic layer, which can be possibly realized by an elastic metamaterial. A realization of the metamaterial with slender, tilted micro voids inserted in an isotropic medium is also illustrated. (b) The L-to-L mode transmission and the L-to-S mode conversion as a function of the frequency $f$ ($d$: layer thickness) for different combinations of $C_{11}$ and $C_{66}$ with the fixed value of $C_{11}C_{66} - C_{16}^2$ in mode-coupled ($C_{16} \neq 0$) layers. The transmission calculations were carried by the exact formula using the transfer matrix approach.

By applying Hooke's law and the linearized strain-displacement relation to Eq. (1), the following relation that involves the scattering matrix $S$ ($S_{ij}$) can be derived for the L-mode incidence:

$$\begin{bmatrix} t_l e^{-ik_l d} \\ 0 \\ t_s e^{-ik_s d} \\ 0 \end{bmatrix} = \begin{bmatrix} S_{11} & S_{12} & S_{13} & S_{14} \\ S_{21} & S_{22} & S_{23} & S_{24} \\ S_{31} & S_{32} & S_{33} & S_{34} \\ S_{41} & S_{42} & S_{43} & S_{44} \end{bmatrix} \begin{bmatrix} 1 \\ r_l \\ 0 \\ r_s \end{bmatrix} = \mathbf{S} \cdot \begin{bmatrix} 1 \\ r_l \\ 0 \\ r_s \end{bmatrix}. \quad (2)$$

This equation can be obtained [23] if the field continuity conditions are applied at the left and right sides of the mode-coupled layer of thickness $d$ inserted in an isotropic medium, as sketched in Fig. 1(a). In Eq. (2), the subscripts $l$ and $s$ represent the L and S modes, respectively. Symbols $t$ ($|t|^2$) and $r$ ($|r|^2$) denote the transmission and reflection coefficients (powers) of the mode-coupled layer, respectively, for the L and S modes. The wave numbers $k_{ql}$ and $k_{qs}$ correspond to the quasi-longitudinal (QL) and quasi-shear (QS) modes of the mode-coupled layer, respectively.

To facilitate theoretical analysis, we assume that the reflection coefficients of the mode-coupled layer are small enough ($|r_l|, |r_s| \ll 1$) and that the L and S modes are weakly coupled in the mode-coupled layer. The latter assumption implies that $C_{16}^2 \ll C_{11} C_{66}$. Indeed, one can show that invoking the latter assumption leads to the former assumption [23].

By the former assumption, Eq. (2) yields $t_l \approx S_{11} e^{ik_l d}$ and $t_s \approx S_{31} e^{ik_s d}$. Following Ref. [23], one can obtain

$$|t_s|^2 \approx |S_{31}|^2 = \tilde{A}^2 + \frac{\tilde{B}^2 + \tilde{C}^2}{2} \\ -(\tilde{A}^2 - \tilde{B}\tilde{C}) \cdot \cos(k_{qs} d - k_{ql} d) + G(k_{ql}, k_{qs}), \quad (3)$$

with

$$G(k_{ql}, k_{qs}) \equiv \frac{\tilde{A}^2 - \tilde{B}^2}{2} \cdot \cos(2 k_{ql} d) - \frac{\tilde{C}^2 - \tilde{A}^2}{2} \cdot \cos(2 k_{qs} d) \\ -(\tilde{A}^2 + \tilde{B}\tilde{C}) \cdot \cos(k_{qs} d + k_{ql} d), \quad (4)$$

The explicit formulae for $\tilde{A}, \tilde{B},$ and $\tilde{C}$ are given in Ref. [23].

If the latter assumption holds, the wiggly behavior of the transmission spectra caused by the coupling-induced instability, also reported in crystalline solids [4,5,24,25], disappears. In this case, the term $G(k_{ql}, k_{qs})$ in Eq. (4) can be neglected (See Ref. [23] for details). Therefore, the normalized L-to-S conversion power, $T_S$, can be expressed as

$$T_S = \sqrt{C_{66}^0 / C_{11}^0} \, |t_s|^2 \approx A_S - B_S \cos(k_{qs}d - k_{ql}d), \tag{5}$$

where the superscript "0" stands for the quantities of an isotropic medium adjacent to the mode-coupled layer. Likewise, the normalized L-to-L transmission power, $T_L$, can be simplified to

$$T_L = |t_l|^2 \approx A_L + B_L \cos(k_{qs}d - k_{ql}d). \tag{6}$$

The amplitudes $A_S$, $B_S$, $A_L$, and $B_L$, explicitly given in Ref. [23], are frequency-independent real-valued functions of $\rho_0$, $\rho$, $C_{ij}^0$, and $C_{ij}$.

Finally, the TFPR condition for the maximum L-to-S mode (or S-to-L mode, similarly) conversion can be derived by using the formula for $T_S$ in Eq. (5). The result is surprisingly simple:

$$\Delta\phi \equiv k_{qs}d - k_{ql}d = \Delta\phi|_{theory} \,;\quad \Delta\phi|_{theory} \equiv (2m+1)\pi \quad (m: \text{integer}) \tag{7}$$

Equation (7) states that if the phase difference $\Delta\phi$ of two dissimilar modes through a mode-coupled anisotropic layer is exactly odd multiples of $\pi$, the so-called TFPR occurs.

On the other hand, $T_L$ becomes maximized if $\Delta\phi = 2m\pi$ (m: integer) as expected from Eq. (6). Accordingly, the number of TFPR (or anti-TFPR) points is fewer than that of UFPR points ($kd = m\pi$) in the same frequency range. This means that the first anti-TFPR frequency can be the same as the first UFPR frequency only if the mode-coupled layer is over twice as thick as a layer of UFPR. Another important observation from Fig. 1(b) is that the transmission spectra are wiggly as $\Delta\phi$ (equivalently, the frequency $f$) varies. This wiggly behavior was due to the presence of the $G$ term in Eq. (4), which was ignored for the theoretical derivation of the TFPR condition (see Ref. [23] for details). We will examine the wiggly behavior by using Fig. 2.

Figure 2 shows the L-to-L and L-to-S transmission spectra of weakly and strongly mode-coupled layers. The degree of the mode coupling is determined by the relative ratio of $C_{16}$ to $C_{11}$ (and $C_{66}$). Actually, the ratio is directly related to the degree of structural instability [4,5,24,25], which can be described in terms of the structural instability factor $F_{SI}$ defined as

$$F_{SI} = \frac{C_{11}^0 \cdot C_{66}^0}{C_{11} \cdot C_{66} - C_{16}^2}. \tag{8}$$

Figure 2 shows that when $F_{SI}$ is small (say, less than 5), the theoretical result (7) accurately predicts the $\Delta\phi|_{max}$ values where $T_S$ is maximized, and the transmission spectra are not wiggly. If $F_{SI}$ is not small (*i.e.*, for strongly-coupled media), there is some discrepancy between $\Delta\phi|_{theory}$ and $\Delta\phi|_{max}$ with significantly wiggly transmission curves. The deviation and the formation of the wiggles were mainly due to the ignorance of the *G* term in Eq. (4). Due to the assumptions used to derive Eq. (7), the transmission values were larger than unity. As shown in Fig. 2, media with larger $F_{SI}$ make the transmission spectra strongly wiggly. In recent studies [26-28], the notion of structural instability has been also used to control wave propagation characteristics. Nevertheless, the theoretical TFPR condition (7) is accurate as long as $F_{SI}$ is not large.

At the first and second TFPRs, the distributions of the particle velocities ($v_x, v_y$) of the QL and QS modes are also shown in Fig. 2. To have higher L-to-S mode conversion rates, the velocity magnitude of the QS mode should have the same order of that of the QL mode. At the same time, the magnitudes of $v_x$ and $v_y$ of both the QL and QS modes should be close as much as possible in order to maximize $T_S$ and to minimize $T_L$. Therefore, the particle-velocity polarization of the QL and QS modes being at $+45°$ and $-45°$, respectively, is needed to have high L-to-S (or S-to-L) mode conversion through the mode-coupled layer. Using transient analysis [23], we also found that the strongly-coupled medium requires a 2- to 3-fold longer time to reach a steady-state TFPR state than the weakly-coupled medium does.

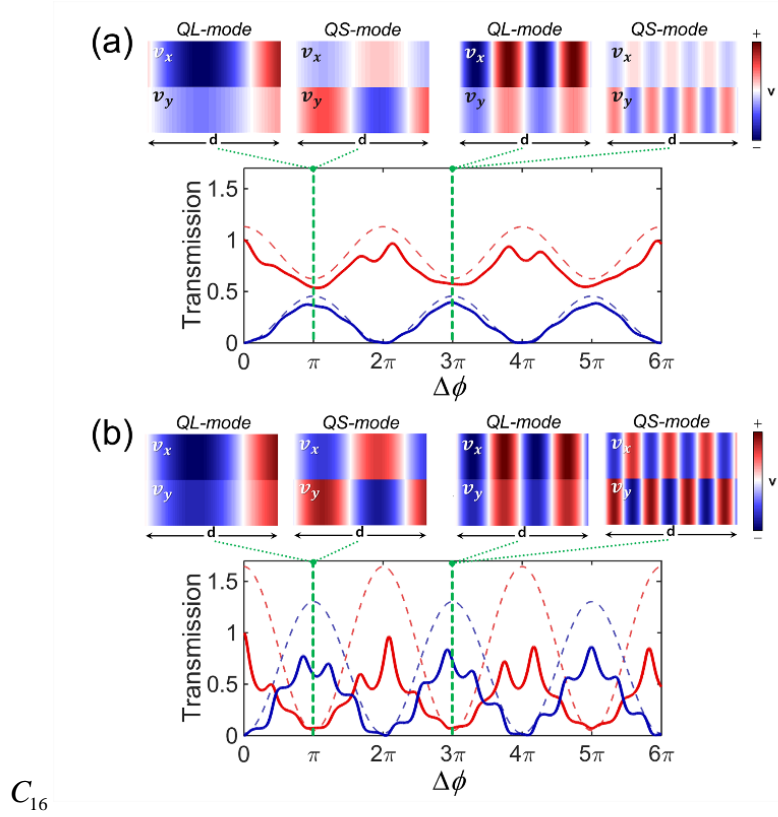

$C_{16}$

FIG. 2 (color online). Transmission spectra as a function of $\Delta\phi = (k_{qs} - k_{ql})d$ for (a) the weakly-coupled layer with $F_{SI}$=3.04 ($C_{11}$=45.2 GPa, $C_{66}$=17.9 GPa, $C_{16}$=10.4 GPa, $\rho$=2390 kg/m$^3$) and (b) the strongly-coupled layer with $F_{SI}$=10.06 ($C_{11}$=20.5 GPa, $C_{66}$=12.9 GPa, =7.22 GPa, $\rho$=2260 kg/m$^3$). The L-mode incidence into the mode-coupled layers is considered. Blue and red curves represent $T_L$ and $T_S$, respectively. Solid lines represent the exact results while the dotted lines, the theoretical results obtained with the assumptions of $|r_l|, |r_s| <<1$ and $C_{16}^2 << C_{11}C_{66}$. The insets show the distribution of the particle velocity components ($v_x, v_y$) of the QL and QS modes at $\Delta\phi|_{theory} = \pi$ and $3\pi$ where $x$ is the direction of an incident L wave.

To see the relationship between the structural instability and the wiggly behavior of the transmission spectra, the contour of $T_S$ is plotted on the $\Delta\phi - F_{SI}$ plane in Fig. 3. To increase the $F_{SI}$ value monotonically, all components of $C_{11}$, $C_{66}$, and $C_{16}$ were linearly varied starting from their nominal values (See the details in Ref. [23]). We found that the transmission spectra become wigglier along the $F_{SI}$ axis with narrower bandwidths. The narrowed bandwidth implies that there was the strong confinement (or long lifetime) of the elastic-wave fields inside the layer at TFPRs. Nevertheless, the highest $T_S$ values for a given $F_{SI}$ were observed near or at $\Delta\phi = \Delta\phi|_{theory} = \pi, 3\pi, 5\pi, \ldots$, as predicted by Eq. (7).

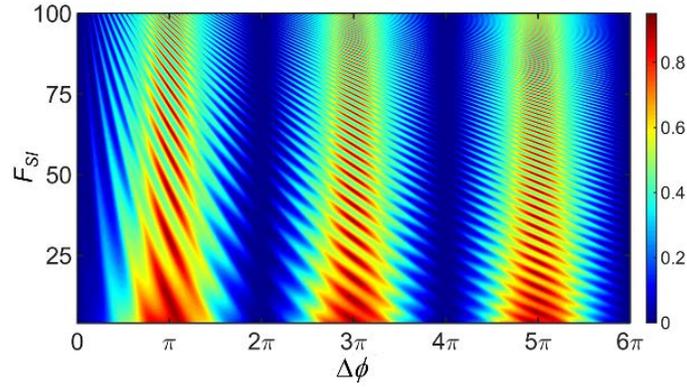

FIG. 3 (color online). Contour of the L-to-S conversion rate $T_S$ on the $\Delta\phi - F_{SI}$ plane. The specific values of $C_{11}$, $C_{66}$, and $C_{16}$ used to determine $F_{SI}$ can be found in Ref. [23].

To validate our findings of the TFPR condition and the wiggly behavior, we performed elastic wave experiments on a 1 mm-thick aluminum plate where the mode-coupled layer made of an elastic metamaterial (EMM) was fabricated by laser beam machining. The engineered EMM is a single-phase EMM the unit cell of which has an oblique void slit. The SEM images of the unit cell configurations are shown in Fig. 4(a). Void slits were found to effectively realize non-resonant type EMMs exhibiting anisotropy needed for elastic wave control [30,31]. EMMs shown Fig. 4(a) were fabricated. The void slits in EMMs must be titled at a certain angle with respect to the wave incidence direction (here, the horizontal direction) for mode conversion through the EMM layer. The weakly and strongly coupled EMM layers consist of eight and six periods of the designed unit cells, respectively.

The wave incident from the surrounding homogeneous Al plate to the EMM layers is an in-plane longitudinal plane wave, specifically, the $S_0$ mode (the lowest symmetric Lamb wave mode) wave as used in Refs. [17,21]. This mode has a predominant longitudinal displacement component in the $x$ direction. To validate our theoretical analysis based on the plane longitudinal wave incidence, a transmitter (a magnetostrictive patch transducer, see Refs. [29-31] for details) was taylor-made to generate a uniform field along the $y$ direction.

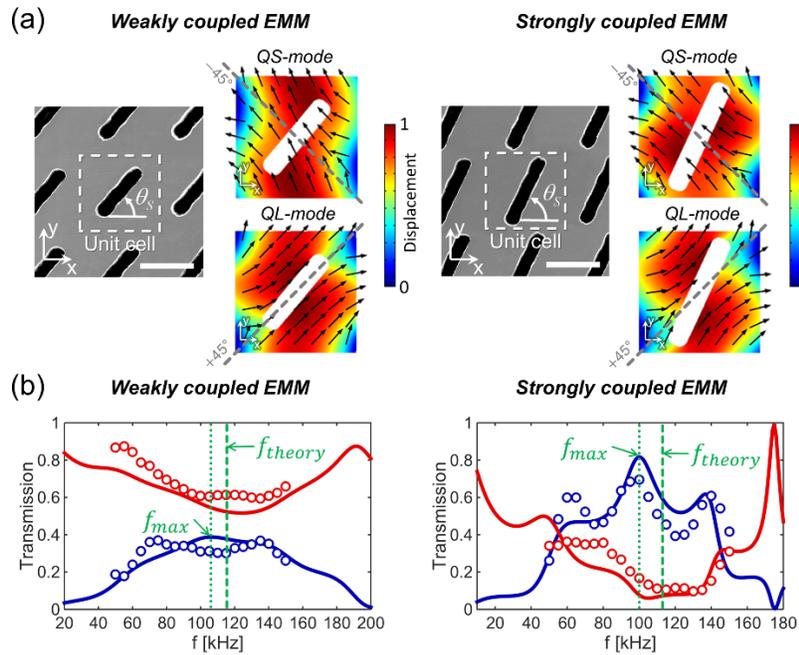

FIG. 4 (color online). (a) SEM images (with 2 mm-long scale bar) and the mode-shapes of $3 \times 3$ mm$^2$ EMM unit cells of weakly mode-coupled EMM (left) and strongly mode-coupled EMM (right) at $f_{max} \approx$ 100 kHz. [Slit length, thickness, rotation angle $\theta_s$] = [2.1 mm, 0.5 mm, 50°] (Weakly coupled) and [2.8 mm, 0.6 mm, 65°] (Strongly coupled). (b) The transmission spectra of $T_S$ (in blue) and $T_L$ (in red) near $f_{theory}$ = 115 kHz for the weakly coupled EMM and near $f_{theory}$ = 113 kHz for the strongly coupled EMM. Solid lines: simulation results, circles: experimental results.

The experimental results for $T_L$ and $T_S$ are marked with circles in Fig. 4(b). The results are in good agreement with the simulation results (solid lines). However, some discrepancy between the numerical and experimental S-mode transmissions results from errors in calibrating the magnitude of the converted $SH_0$ mode with respect to the incident $S_0$ mode. Because the coupling between the transducer and the test specimen cannot be perfect, some errors can be also introduced during measurements (see Ref. [23] for details).

The simulation results in Fig. 4(b) were obtained by using finite element calculations based on fully-detailed unit cell models [23]. The S-parameter method [32] was used to retrieve $C_{ij}$ for the unit cells shown in Fig. 4(a), and then the retrieved $C_{ij}$'s were used to calculate $f_{theory}$. The theoretical TFPR frequency $f_{theory}$ predicted by Eq. (7) is reasonably close to $f_{max}$ (the frequency where $T_S$ is maximized): $f_{theory}=106$ kHz and $f_{max}=115$ kHz for the weakly coupled EMM, and $f_{theory}=100$ kHz and $f_{max}=113$ kHz for the strongly coupled EMM. The discrepancy between $f_{theory}$ and $f_{max}$ is due to inaccuracy in retrieving $C_{ij}$'s and the violation of the weakly-coupled assumption. The error in the retrieval may not be unavoidable; in the wide frequency range, there is a trade-off between the size of unit cells and the EMM layer thickness (*i.e.*, the number of unit cells). In Fig. 4(a), the arrows denote the direction of the displacement field, and the color levels correspond to the displacement magnitude. From the field distributions, it is clear that the displacement-field directions of both the QL and QS modes need to make $\pm 45°$ with respect to the *x* axis (wave incidence direction) to achieve high mode-conversion rates. Even though not presented here, one can show that high S-to-L mode conversion occurs at the same TFPRs.

In summary, the phenomenon of the maximum mode-conversion, from longitudinal to shear modes, in elastic mode-coupled layers is discovered and explained by the TFPR condition. At the TFPR frequency, the phase difference between the QL and QS modes generally satisfies odd multiples of π. High mode conversion rates can be achieved through mode-coupled layers exhibiting the strong QS mode confinement and near $\pm 45°$ polarization angles of both the QL and QS modes. The transmission spectra of mode-coupled layers can be wiggly for which the structural instability is responsible. We realized mode-coupled media by designing slit-type EMMs which have wide programmability of the conversion rate and band flatness. The TFPR of the mode-coupled EMMs was successfully observed by using simulations and ultrasonic experiments. Our results provide the fundamental understanding of elastic-wave mode conversion in anisotropic media, which has a great potential in shear-wave based ultrasonic flaw-inspection, transmission control at fluid/solid interfaces [9,10,15], and maximized wave dissipation in viscoelastic media [12-14] for many industrial and biomedical applications.

This research was supported by the Global Frontier R&D Program on Center for Wave Energy Control based on Metamaterials funded by the Korean Ministry of Science, ICT & Future Planning, contracted through Institute of Advanced Machines and Design at Seoul National University in Korea (2014M3A6B3063711).

[*]Corresponding author: yykim@snu.ac.kr

# Supplemental Material for
# "Transmodal Fabry-Pérot Resonance: Theory and Realization with Elastic Metamaterials"

Joshua Minwoo Kweun,[1] Hyung Jin Lee,[2] Joo Hwan Oh,[3] Hong Min Seung,[4] and Yoon Young Kim[1,2,*]

[1] *School of Mechanical and Aerospace Engineering, Seoul National University, 1 Gwanak-ro, Gwanak-gu, Seoul 08826, South Korea*
[2] *Institute of Advanced Machines and Design, Seoul National University, 1 Gwanak-ro, Gwanak-gu, Seoul 08826, South Korea*
[3] *School of Mechanical and Nuclear Engineering, Ulsan National Institute of Science and Technology, UNIST-gil 50, Eonyang-eup, Ulju-gun, Ulsan 44919, South Korea*
[4] *Center for Safety Measurement, Division of Metrology for Quality of Life, Korea Research Institute of Standards and Science (KRISS), 267 Gajeong-ro, Yuseong-gu, Daejeon 34113, South Korea*

## THEORETICAL DETAILS

### 1. Theory of the Transmodal Fabry-Pérot Resonance

We provide the derivation procedure of the transmodal Fabry-Pérot resonance (TFPR) condition in detail. We consider harmonic plane waves on the *xy*-plane in elastic, anisotropic, and mode-coupled media. If a plane wave travels along the *x* direction, the displacement field **u** in elastic media can be written as

$$\mathbf{u} = \left( A\mathbf{P}_{ql}e^{+ik_{ql}x} + B\mathbf{P}_{ql}e^{-ik_{ql}x} + C\mathbf{P}_{qs}e^{+ik_{qs}x} + D\mathbf{P}_{qs}e^{-ik_{qs}x} \right)e^{-i\omega t}, \quad (S1)$$

where $A, B, C,$ and $D$ are the displacement amplitudes of the quasi-longitudinal (QL) and quasi-shear (QS) modes. The polarization vectors of the QL and QS-modes are denoted by $\mathbf{P}_{ql} = P_x^{ql}\hat{\mathbf{x}} + P_y^{ql}\hat{\mathbf{y}}$ and $\mathbf{P}_{qs} = P_x^{qs}\hat{\mathbf{x}} + P_y^{qs}\hat{\mathbf{y}}$, respectively. The symbols $k_{ql}$ and $k_{qs}$ represent the wave numbers of the QL and QS modes, respectively. To find the polarization vectors (corresponding to eigenvectors of the system) and wave numbers (eigenvalues), we solve the following *Christoffel* equation:

$$k^2 \mathbf{\Gamma} \mathbf{v} = k^2 \begin{bmatrix} C_{11} & C_{16} \\ C_{16} & C_{66} \end{bmatrix} \begin{bmatrix} v_x \\ v_y \end{bmatrix} = \rho \omega^2 \begin{bmatrix} v_x \\ v_y \end{bmatrix}, \quad (S2)$$

where $\mathbf{\Gamma}$ is the *Christoffel* matrix, and $\mathbf{v}$ is the polarization vector, and $C_{ij}$'s and $\rho$ represent the stiffness constants and mass density of the anisotropic mode-coupled medium, respectively. For a nontrivial solution, the characteristic determinant of Eq. (S2) should be zero as

$$\Omega(\omega, k) = \left| k^2 \Gamma_{ij} - \rho \omega^2 \delta_{ij} \right| = 0, \quad (S3)$$

Solving Eq. (S3), we get the expressions for the wave number k and the polarization vectors **P** as (upper and lower signs for the QL and QS modes, respectively)

$$k^2 = \frac{\rho \omega^2 \left\{ C_{11} + C_{66} \mp \sqrt{(C_{11} - C_{66})^2 + 4C_{16}^2} \right\}}{2(C_{11}C_{66} - C_{16}^2)}, \quad (S4)$$

and

$$P_x = \frac{C_{16}}{|C_{16}|} \left[ \frac{1}{2} \cdot \left\{ Q \pm \left( \frac{C_{66} - C_{11}}{|C_{16}|} \right) \sqrt{Q} \right\} \right]^{-1/2}, \quad P_y = \pm \left[ 1 + 4 \cdot \left( \frac{C_{66} - C_{11}}{|C_{16}|} \pm \sqrt{Q} \right)^{-2} \right]^{-1/2}, \quad Q = 4 + \left( \frac{C_{66} - C_{11}}{C_{16}} \right)^2. \quad (S5)$$

For normal wave incidence, it is convenient to express the field variables (velocity components $v_x$ and $v_y$ of the velocity vector $\mathbf{v} = -i\omega \mathbf{u}$; stress components $\sigma_{xx}$ and $\sigma_{xy}$) in terms of the displacement amplitude variables as

$$\mathbf{f}_x \equiv \begin{bmatrix} v_x \\ v_y \\ \sigma_{xx} \\ \sigma_{xy} \end{bmatrix} = \mathbf{M} \mathbf{N}\big|_x \begin{bmatrix} A \\ B \\ C \\ D \end{bmatrix}, \quad (S6)$$

where the matrix **M** is given as

$$\mathbf{M} = \begin{bmatrix} -j\omega P_x^{ql} & -j\omega P_x^{ql} & -j\omega P_x^{qs} & -j\omega P_x^{qs} \\ -j\omega P_y^{ql} & -j\omega P_y^{ql} & -j\omega P_y^{qs} & -j\omega P_y^{qs} \\ jk_{ql}(C_{11}P_x^{ql} + C_{16}P_y^{ql}) & -jk_{ql}(C_{11}P_x^{ql} + C_{16}P_y^{ql}) & jk_{qs}(C_{11}P_x^{qs} + C_{16}P_y^{qs}) & -jk_{qs}(C_{11}P_x^{qs} + C_{16}P_y^{qs}) \\ jk_{ql}(C_{16}P_x^{ql} + C_{66}P_y^{ql}) & -jk_{ql}(C_{16}P_x^{ql} + C_{66}P_y^{ql}) & jk_{qs}(C_{16}P_x^{qs} + C_{66}P_y^{qs}) & -jk_{qs}(C_{16}P_x^{qs} + C_{66}P_y^{qs}) \end{bmatrix}, \quad (S7)$$

and the matrix $\mathbf{N}\big|_x$ containing the spatial phase information at position *x* is written as

$$\mathbf{N}\big|_x = \begin{bmatrix} e^{jk_{ql}x} & 0 & 0 & 0 \\ 0 & e^{-jk_{ql}x} & 0 & 0 \\ 0 & 0 & e^{jk_{qs}x} & 0 \\ 0 & 0 & 0 & e^{-jk_{qs}x} \end{bmatrix}. \tag{S8}$$

In deriving Eqs. (S6-S8), we used the elastic constitutive equation (or Hooke's law) $\boldsymbol{\sigma} = \mathbf{C}:\boldsymbol{\varepsilon}$, where $\boldsymbol{\sigma}$ is the stress field and $\boldsymbol{\varepsilon} = \tfrac{1}{2}(\nabla\mathbf{u} + \nabla\mathbf{u}^T)$ is the strain field.

From the field-amplitude relationship in Eq. (S6), the velocity fields of the QL and QS modes can be easily calculated. Their real parts are written as

$$\begin{aligned}
\mathrm{Re}(\mathbf{v}_{ql}) &= \omega \cdot \mathbf{P}_{ql} \cdot \big[ \mathrm{Im}(A+B) \cdot \cos(k_{ql}x) - \mathrm{Re}(A-B) \cdot \sin(k_{ql}x) \big], \\
\mathrm{Re}(\mathbf{v}_{qs}) &= \omega \cdot \mathbf{P}_{qs} \cdot \big[ \mathrm{Im}(C+D) \cdot \cos(k_{qs}x) - \mathrm{Re}(C-D) \cdot \sin(k_{qs}x) \big].
\end{aligned} \tag{S9}$$

Using the expressions in Eqs. (S7) and (S8), the transfer matrix $\mathbf{T}$ relating the field variables $\mathbf{f}_x$ at $x$ and $\mathbf{f}_{x+d}$ at $x+d$ can be found as, where $d$ is the thickness of the elastic mode-coupled layer in consideration,

$$\mathbf{f}_{x+d} = \mathbf{T} \cdot \mathbf{f}_x, \qquad \mathbf{T} \equiv \mathbf{M}\mathbf{N}\big|_{x=d}\,\mathbf{M}^{-1}. \tag{S10}$$

Now we will apply the continuity conditions of the elastic fields at the interfaces of the mode-coupled layer surrounded by an isotropic medium:

$$\mathbf{f}_{x=0^-} = \mathbf{f}_{x=0^+}, \qquad \mathbf{f}_{x=d^-} = \mathbf{f}_{x=d^+}, \tag{S11}$$

where $x=0^-$ and $x=d^+$ represent the locations of the isotropic medium adjacent to the left ($x=0^+$) and the right ($x=d^-$) sides of the mode-coupled layer, respectively. Then, the $4 \times 4$ scattering matrix $\mathbf{S}$ of the mode-coupled monolayer system is given as

$$\mathbf{S} \equiv \mathbf{M}_0^{-1}\,\mathbf{T}\mathbf{M}_0, \tag{S12}$$

where the matrix $\mathbf{M}_0$ of the isotropic medium is

$$\mathbf{M}_0 = \begin{bmatrix} -j\omega & -j\omega & 0 & 0 \\ 0 & 0 & -j\omega & -j\omega \\ jk_l C_{11}^0 & -jk_l C_{11}^0 & 0 & 0 \\ 0 & 0 & jk_s C_{66}^0 & -jk_s C_{66}^0 \end{bmatrix} \tag{S13}$$

with the pure-longitudinal (L) wave number $k_l$ and the pure-shear (S) wave number $k_s$, and the stiffness constants of the isotropic medium, $C_{11}^0$ and $C_{66}^0$. If an L mode is assumed to be incident upon the mode-coupled layer, the following relation holds as

$$\begin{bmatrix} t_l e^{-ik_l d} \\ 0 \\ t_s e^{-ik_s d} \\ 0 \end{bmatrix} = \mathbf{S} \cdot \begin{bmatrix} 1 \\ r_l \\ 0 \\ r_s \end{bmatrix} = \begin{bmatrix} S_{11} & S_{12} & S_{13} & S_{14} \\ S_{21} & S_{22} & S_{23} & S_{24} \\ S_{31} & S_{32} & S_{33} & S_{34} \\ S_{41} & S_{42} & S_{43} & S_{44} \end{bmatrix} \begin{bmatrix} 1 \\ r_l \\ 0 \\ r_s \end{bmatrix}, \tag{S14}$$

where $t_l$ and $t_s$ are the transmission coefficients of the L and S modes, respectively, transmitted into the isotropic medium next to the right interface between the isotropic and anisotropic mode-coupled media, and $r_l$ and $r_s$ are the reflection coefficients of the L and S modes at the left interface, respectively.

To derive an analytic expression for the condition of the transmodal Fabry-Pérot resonance, we assume that the reflection of the mode-coupled layer is small enough ($|r_l|, |r_s| \ll 1$). This condition is related to the condition that the structural instability factor $F_{SI}$ [defined in Eq. (8) or Eq. (S26)] is small. With this assumption, the transmission coefficients can be simplified as

$$\begin{aligned}
t_l e^{-ik_l d} &= S_{11} + S_{12} \cdot r_l + S_{14} \cdot r_s \approx S_{11}, \\
t_s e^{-ik_s d} &= S_{31} + S_{32} \cdot r_l + S_{34} \cdot r_s \approx S_{31}.
\end{aligned} \tag{S15}$$

Thus, the magnitude of the transmission coefficient $t_s$ of the converted S mode can be written as

$$|t_s|^2 \approx |S_{31}|^2 = \tilde{A}^2 + \frac{\tilde{B}^2 + \tilde{C}^2}{2} - \left(\tilde{A}^2 - \tilde{B}\tilde{C}\right)\cos(k_{qs}d - k_{ql}d) + G(k_{ql}, k_{qs}), \tag{S16}$$

where

$$G(k_{ql}, k_{qs}) \equiv \frac{\tilde{A}^2 - \tilde{B}^2}{2} \cdot \cos(2k_{ql}d) - \frac{\tilde{C}^2 - \tilde{A}^2}{2} \cdot \cos(2k_{qs}d) - \left(\tilde{A}^2 + \tilde{B}\tilde{C}\right)\cos(k_{qs}d + k_{ql}d). \tag{S17}$$

If the coupling constant $C_{16}$ is positive, we get

$$\tilde{A} = -\frac{P_y^{ql} P_y^{qs}}{2} - \frac{k_l C_{11}^0}{2k_s C_{66}^0 \left(C_{11}C_{66} - C_{16}^2\right)} \left(C_{16} P_x^{ql} + C_{66} P_y^{ql}\right)\left(C_{16} P_x^{qs} + C_{66} P_y^{qs}\right),$$

$$\tilde{B} = -\frac{k_{ql} P_y^{qs}}{2k_s C_{66}^0} \left(C_{16} P_x^{ql} + C_{66} P_y^{ql}\right) - \frac{k_l C_{11}^0 P_y^{ql}}{2k_{ql} \left(C_{11}C_{66} - C_{16}^2\right)} \left(C_{16} P_x^{qs} + C_{66} P_y^{qs}\right), \quad \text{(S18)}$$

$$\tilde{C} = \frac{k_{qs} P_y^{ql}}{2k_s C_{66}^0} \left(C_{16} P_x^{qs} + C_{66} P_y^{qs}\right) + \frac{k_l C_{11}^0 P_y^{qs}}{2k_{qs} \left(C_{11}C_{66} - C_{16}^2\right)} \left(C_{16} P_x^{ql} + C_{66} P_y^{ql}\right).$$

The quantities with the superscript "0" are those associated with the isotropic medium surrounding the mode-coupled layer. If $C_{16}$ is negative, the sign of all values $\tilde{A}$, $\tilde{B}$, and $\tilde{C}$ becomes opposite.

To simplify the above expressions further, we assume $C_{11}C_{66} \gg C_{16}^2$ which implies that the longitudinal and shear motions of the mode-coupled layer are weakly coupled. This weak-coupling condition can be satisfied if $C_{ij} \approx C_{ij}^0$. Therefore, small nonzero $C_{16}$ terms needed for mode-coupled layers can be produced if slender and tilted micro-voids are created in the adjacent isotropic material of $C_{ij}^0$. If this assumption is used, the following approximations can be made

$$\tilde{A} \approx \tilde{B} \approx -\tilde{C} \quad \text{(S19)}$$

The use of Eq. (S19) makes the $G$ term in Eq. (S17) vanish. Accordingly, the wiggly behavior of the transmission spectra caused by the coupling-induced instability can be diminished. Then, Eq. (S16) reduces to

$$|t_s|^2 \approx \tilde{A}^2 + \frac{\tilde{B}^2 + \tilde{C}^2}{2} - \left(\tilde{A}^2 - \tilde{B}\tilde{C}\right) \cdot \cos\left(k_{qs}d - k_{ql}d\right). \quad \text{(S20)}$$

Under the L-mode incidence, the normalized longitudinal-to-shear mode (L-to-S) transmission power ($T_S$), which we also call the L-to-S mode conversion rate, can be approximated as

$$T_S = \sqrt{\frac{C_{66}^0}{C_{11}^0}} |t_s|^2 \approx A_S - B_S \cos\left(k_{qs}d - k_{ql}d\right), \quad A_S = \sqrt{\frac{C_{66}^0}{C_{11}^0}} \left(\tilde{A}^2 + \frac{\tilde{B}^2 + \tilde{C}^2}{2}\right), \quad B_S = \sqrt{\frac{C_{66}^0}{C_{11}^0}} \left(\tilde{A}^2 - \tilde{B}\tilde{C}\right). \quad \text{(S21)}$$

Similarly, the normalized longitudinal-to-longitudinal mode (L-to-L) transmission power ($T_L$) can be simplified to

$$T_L = |t_l|^2 \approx |S_{11}|^2 \approx A_L + B_L \cos\left(k_{qs}d - k_{ql}d\right), \quad A_L = \frac{\hat{A}^2 + \hat{B}^2 + \hat{C}^2 + \hat{D}^2}{2}, \quad B_L = \hat{A}\hat{B} + \hat{C}\hat{D}, \quad \text{(S22)}$$

with

$$\hat{A} \approx \hat{C} \quad \text{and} \quad \hat{D} \approx \hat{B}. \quad \text{(S23)}$$

If the coupling constant $C_{16}$ is positive, we get

$$\hat{A} = -\frac{P_x^{ql} P_y^{qs}}{2} - \frac{\left(C_{11} P_x^{ql} + C_{16} P_y^{ql}\right)\left(C_{16} P_x^{qs} + C_{66} P_y^{qs}\right)}{2\left(C_{11}C_{66} - C_{16}^2\right)},$$

$$\hat{B} = \frac{P_y^{ql} P_x^{qs}}{2} + \frac{\left(C_{11} P_x^{qs} + C_{16} P_y^{qs}\right)\left(C_{16} P_x^{ql} + C_{66} P_y^{ql}\right)}{2\left(C_{11}C_{66} - C_{16}^2\right)},$$

$$\hat{C} = -\frac{k_{ql} P_y^{qs}}{2k_l C_{11}^0} \left(C_{11} P_x^{ql} + C_{16} P_y^{ql}\right) - \frac{k_l C_{11}^0 P_x^{ql}}{2k_{ql} \left(C_{11}C_{66} - C_{16}^2\right)} \left(C_{16} P_x^{qs} + C_{66} P_y^{qs}\right), \quad \text{(S24)}$$

$$\hat{D} = \frac{k_{qs} P_y^{ql}}{2k_l C_{11}^0} \left(C_{11} P_x^{qs} + C_{16} P_y^{qs}\right) + \frac{k_l C_{11}^0 P_x^{qs}}{2k_{qs} \left(C_{11}C_{66} - C_{16}^2\right)} \left(C_{16} P_x^{ql} + C_{66} P_y^{ql}\right).$$

Similarly, if $C_{16}$ is negative, the sign of all values $\hat{A}$, $\hat{B}$, $\hat{C}$, and $\hat{D}$ becomes opposite. Thus, from the Eqs. (S21) and (S22), the TFPR condition can be obtained as

$$\Delta k = k_{qs} - k_{ql} = \frac{(2m+1)\pi}{d} \quad (m: \text{integer}), \quad \text{(S25a)}$$

which is equivalent to

$$\Delta \phi \equiv \Delta k d = \Delta \phi |_{theory} \quad \text{with} \quad \Delta \phi |_{theory} \equiv (2m+1)\pi \quad (m: \text{integer}) \quad \text{(S25b)}$$

Equation (S25) is the same as Eq. (7).

## 2. The effects of structural instability on wiggly transmission responses through mode-coupled layers

In this section, we examine the effect of the structural instability on the wiggly behavior of the transmission spectra through mode-coupled layers. The structural instability defined as Eq. (8) is rewritten here as Eq. (S26) for the sake of convenience:

$$F_{SI} = \frac{C_{11}^0 C_{66}^0}{C_{11}C_{66} - C_{16}^2}. \quad \text{(S26)}$$

Because the contour plot of $T_S$ in the $\Delta\phi - F_{SI}$ plane is given as Fig. 3, we here show the contour plots of the L-to-L transmission $T_L$, the term $G(k_{ql}, k_{qs})$, and the reflectance $R$ (Fig. S1). The reflectance is defined as below:

$$R = |r_l|^2 + \sqrt{\frac{C_{66}^0}{C_{11}^0}} |r_s|^2. \tag{S27}$$

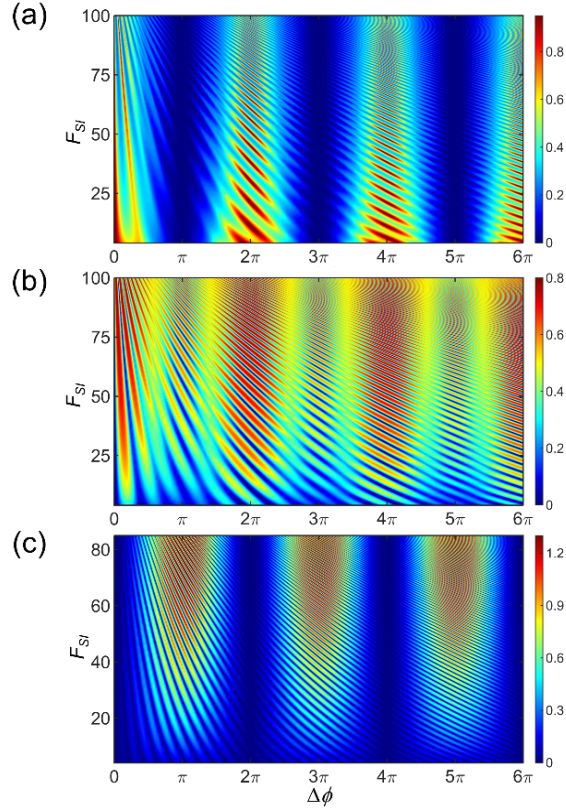

FIG. S1. Contours of (a) the L-to-L transmission $T_L$, (b) the reflectance $R$ of the mode-coupled layer, and (c) the term $G(k_{ql}, k_{qs})$ on the $\Delta\phi - F_{SI}$ plane.

In Fig. 3 and also in Fig. S1, specific combinations of ($C_{11}, C_{66}, C_{16}$) values were used. Here, we simply chose linearly varying $C_{ij}$'s as

$$\frac{C_{11} - 28.97\,\text{GPa}}{-24.05\,\text{GPa}} = \frac{C_{66} - 26.92\,\text{GPa}}{-22.92\,\text{GPa}} = \frac{C_{16} - 12.39\,\text{GPa}}{-12.48\,\text{GPa}} \tag{S28}$$

with $\rho = 2700$ kg/m$^3$ and $fd = 3$ MHz-mm. Actually, the selected $C_{ij}$ variation is indicated as a white line connecting $P$ and $Q$ in the color contour of $F_{SI}$ for varying $C_{11}$, $C_{66}$, and $C_{16}$ in Fig. S2.

Figure S1 shows that the maximum L-to-S conversion and the maximum L-to-L transmission occur at $\Delta\phi \approx (2m+1)\pi$ and $\Delta\phi \approx (2m)\pi$ ($m$: integer), respectively. In addition, the behavior of $F_{SI}$ is strongly correlated with $R$ and $G(k_{ql}, k_{qs})$. This implies that the larger $R$ and $G(k_{ql}, k_{qs})$ are, the wigglier the transmission spectra are. Recall that we argued in the main part of the paper that smaller $F_{SI}$ values correspond to smaller reflection coefficients (*i.e.*, smaller $R$ values). To show this, Fig. S3 is prepared.

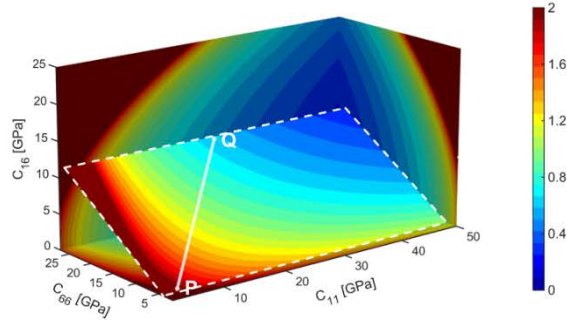

FIG. S2. Color contour of the structural instability factor $F_{SI}$ (in log scale) for varying $C_{11}$, $C_{66}$, and $C_{16}$. The line connecting $P$ and $Q$ represents the locus of linearly-varying $F_{SI}$'s. The cutting plane is represented by the dashed lines for easy viewing.

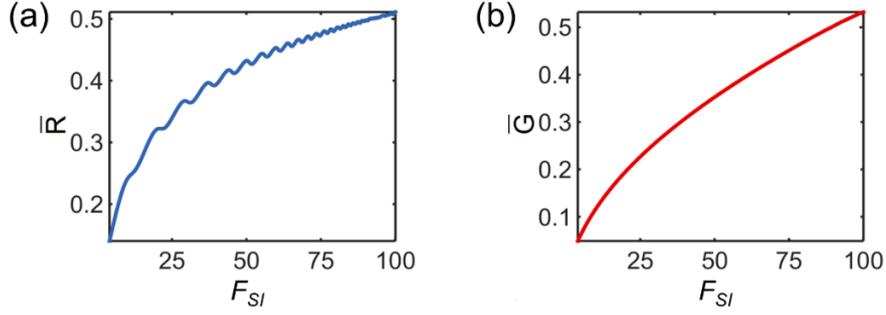

FIG. S3. (a) The mean reflectance $\bar{R}$ and (b) the mean $G$ term $\bar{G}$ as a function of $F_{SI}$.

In Fig. S3, the mean reflectance $\bar{R}$ and the mean $G$ term $\bar{G}$, which are defined below, are plotted as a function of $F_{SI}$:

$$\bar{R} = \frac{1}{2\pi} \int_0^{2\pi} R \, d(\Delta\phi),$$
$$\bar{G} = \frac{1}{2\pi} \int_0^{2\pi} G(k_{ql}, k_{qs}) \, d(\Delta\phi).$$
(S29)

As Fig. S3(a) shows, $\bar{R}$ increases almost monotonically as $F_{SI}$ increases except some oscillations of small magnitude. From Fig. S3(b), $F_{SI}$ is shown to be closely correlated with $\bar{G}$, implying that the structural instability also governs the $G$ term. Therefore, unless $F_{SI}$ is small, the reflectance $R$ and the $G$ term cannot be ignored because the transmission spectra exhibit significantly wiggly behavior.

## SIMULATION DETAILS

We proposed elastic metamaterials (EMMs) having slit-type microstructures to realize anisotropic mode-coupled layers. To calculate the theoretical TFPR frequency $f_{theory}$ from the fabricated EMMs, the effective parameters ($\rho, C_{11}, C_{66}, C_{16}$) of the EMMs should be determined. To this end, we used the S-parameter retrieval method [1, 2] developed for elastic media. The method uses the time-harmonic full finite element simulations (COMSOL Multiphysics), as shown in Fig. S4. As before, the L-mode wave propagating along the horizontal direction is assumed to be incident onto the mode-coupled EMM layer. The selected frequency for the simulation results in Fig. S4 is 100 kHz. Note that the longitudinal and shear wave modes have the dominant $u_x$ ($x$-directional displacement) and $u_y$ ($y$-directional displacement) fields, respectively, when they propagate along the $x$-axis (the horizontal direction). Using the simulation results shown in Fig. S4 and the retrieval method [1, 2], the following values were estimated (at 100 kHz):

$$\rho = 2322 \text{ kg/m}^3, \quad C_{11} = 37.57 \text{ GPa}, \quad C_{66} = 16.28 \text{ GPa}, \quad C_{16} = 8.707 \text{ GPa for weakly-coupled EMM} \quad \text{(S30.a)}$$

$$\rho = 2305 \text{ kg/m}^3, \quad C_{11} = 16.86 \text{ GPa}, \quad C_{66} = 12.85 \text{ GPa}, \quad C_{16} = 7.407 \text{ GPa for strongly-coupled EMM} \quad \text{(S30.b)}$$

The retrieved material parameters given by Eq. (S30) were used to obtain $f_{theory}$ in Fig. 4(b).

While the results in Fig. S4 were used for the retrieval of the effective material parameters, they also provide useful information. After the incident L-mode waves pass through the mode-coupled EMM layers, two distinct wave modes, the L-mode (with dominant $u_x$) and S-mode (with dominant $u_y$), appear on the right side of the homogeneous aluminum plate. Also, the relative magnitude ratio of $u_y/u_x$ becomes larger in the strongly-coupled EMM layer than in the weakly-coupled EMM layer. This means the L-to-S mode conversion rate for the strongly-coupled layer is larger than that for the weakly-coupled layer; this finding is consistent with the results shown by the transmission curves in Fig. 4(b).

Figure 4(b) suggests that the selected frequency of 100 kHz is the frequency of the maximum L-to-S mode conversion for the weakly and strongly coupled EMM layer or a frequency close enough to the frequency. Therefore, this frequency corresponds to a TFPR frequency. To see how transient waves behave after they pass through the mode-coupled EMMs, transient finite element simulations were conducted by employing the same models used to obtain the results in Fig. S4. Ten cycles of L-mode sinusoidal waves centered at 100 kHz were exited at a distance 20 cm away from the left side of the mode-coupled EMMs and the transmitted waves are measured at a distance 10 cm away from the right side of the EMMs. The measured time signals of the normal stress component $\sigma_{xx}$ representing the L mode and shear stress component $\sigma_{xy}$ representing the S mode are shown in Fig. S5. It shows that the L mode wave reaches a steady state a few cycles faster than the S mode wave. It also shows that the strongly-coupled EMM takes a longer time to reach its steady-state TFPR (see the transient signals in the region marked in green) than the weakly-coupled EMM does. This implies that the strongly-coupled EMM has stronger field confinement at the TFPR than the weakly-coupled EMM does.

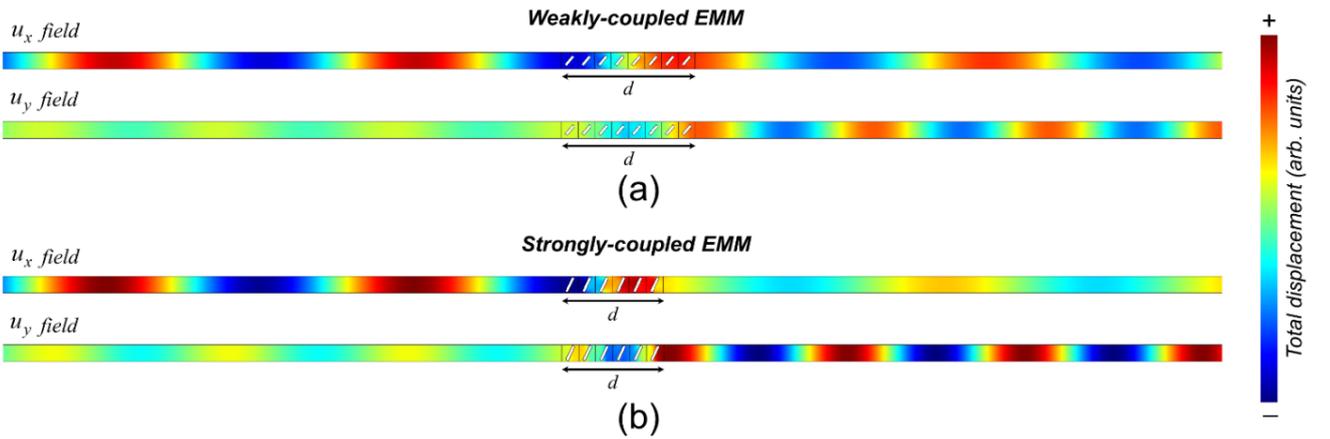

FIG. S4. Time-harmonic wave simulations for longitudinal wave incidence onto mode-coupled EMM layers at 100 kHz. For (a) a weakly-coupled EMM layer (the slab thickness $d$ = 2.4 cm) and (b) a strongly-coupled EMM layer (the slab thickness $d$ = 1.8 cm). The incident wave is assumed to propagate horizontally to the right. The slit length $l$ and thickness $t$ are 2.1 mm and 0.5 mm, respectively, for the weakly-coupled EMM and are 2.8 mm and 0.6 mm, respectively, for the strongly-coupled EMM. The counterclockwise rotation angle $\theta$ of EMM slits is 50° and 65° for the weakly-coupled EMM and strongly-coupled EMM, respectively, with respect to the horizontal line. The side length of each EMM square unit cell is 3 mm. The periodic boundary condition is used in the vertical direction. The perfectly-matched layer condition is also applied in the horizontal direction to absorb transmitted and reflected waves.

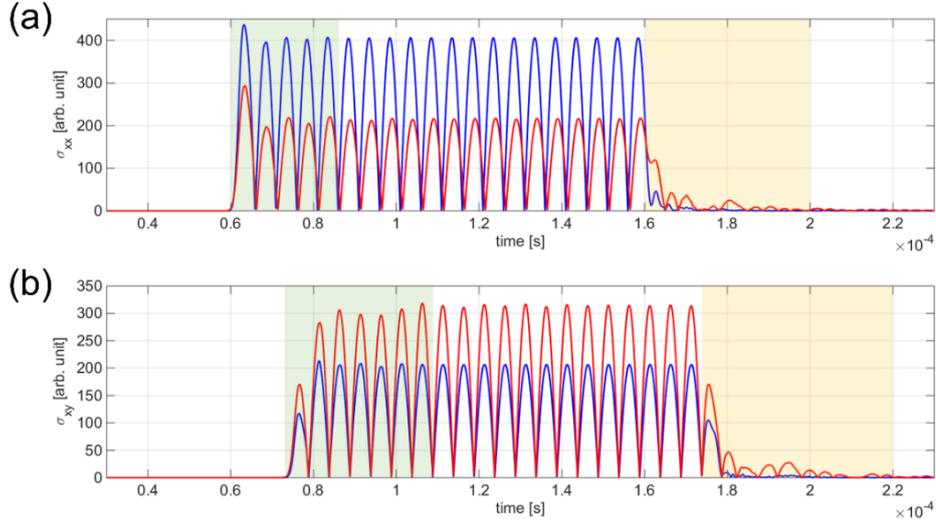

FIG. S5. Transient signals measured at a distance 10 cm from the right side of the mode-coupled EMMs. The same finite element models as those used in Fig. S4 were used. The magnitude of (a) the normal stress $\sigma_{xx}$ that represents the signal of the L-mode wave and (b) the shear stress $\sigma_{xy}$ as the signal of the S-mode wave. Blue lines: for the weakly-coupled EMM (shown in Fig. 4(a)), red lines: for the strongly-coupled EMM (shown in Fig. 4(a)).

The advantage of using EMMs is in programmability. Figure S6 shows that a wide range of the $F_{SI}$ values and the mean L-to-S mode conversion rate $\bar{T}_S$ can be reached by varying the parameters $l$, $t$, and $\theta$ of the slit-type EMMs, as shown in Fig. S6; among the three unit cells shown in Fig. S6(a), the unit cells A, B, and C yield the minimum $F_{SI}$, maximum $\bar{T}_S$, and maximum $F_{SI}$ values, respectively. Figure S7 also illustrates this aspect clearly; it shows the contour plot of $T_S$ by choosing only one parameter among ($l$, $t$, and $\theta$) with varying $\Delta\phi$'s. The silts with large $l$, $t$, and $\theta$ values yield high L-to-S mode conversion rates but narrower bandwidths. These results suggest that, by using the slit-type EMMs, the mode conversion responses can be programmed with a certain degree of freedom. The definition of $\bar{T}_S$ is

$$\bar{T}_S = \frac{1}{2\pi}\int_0^{2\pi} T_S \, d(\Delta\phi) \tag{S31}$$

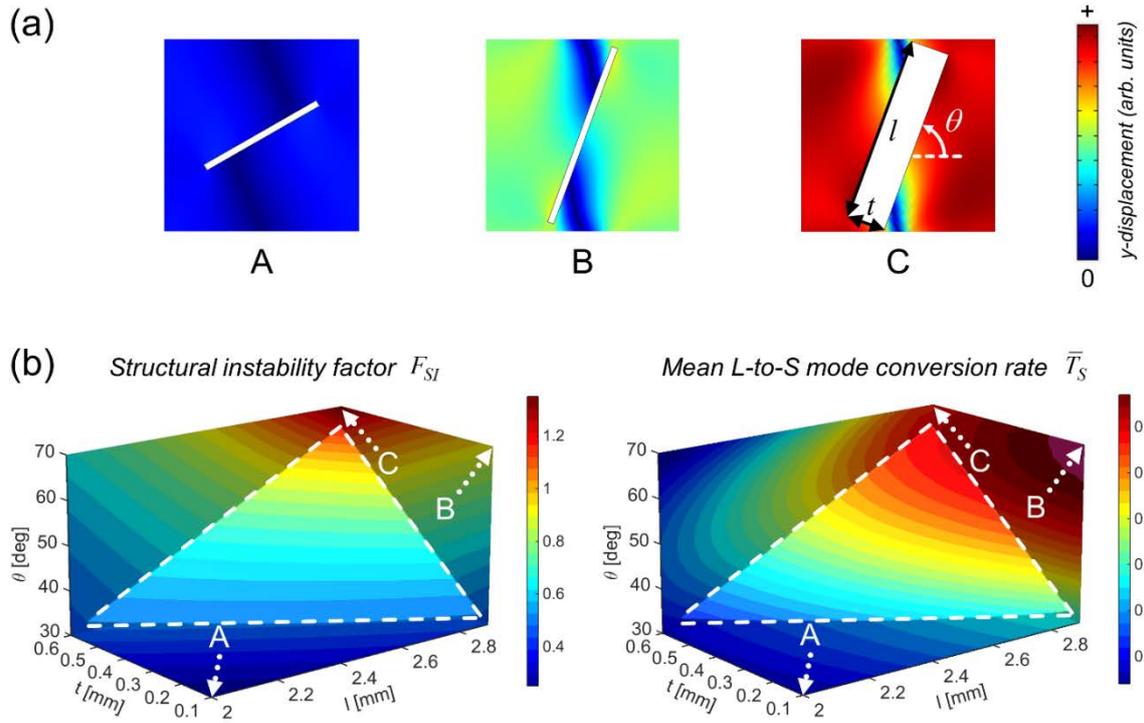

FIG. S6. Programmability with the proposed EMMs. (a) Illustration of selected three EMM unit cells with the color plots of the $|u_y|$ distribution. $(l,t,\theta)$=(2 mm, 0.1 mm, 30°) for unit cell A, (2.9 mm, 0.1 mm, 70°) for unit cell B, and (2.9 mm, 0.6 mm, 70°) for unit cell C. (b) Contour plots of the structural instability $F_{SI}$ (in log$_{10}$ scale) and the mean L-to-S conversion rate $\bar{T}_S$.

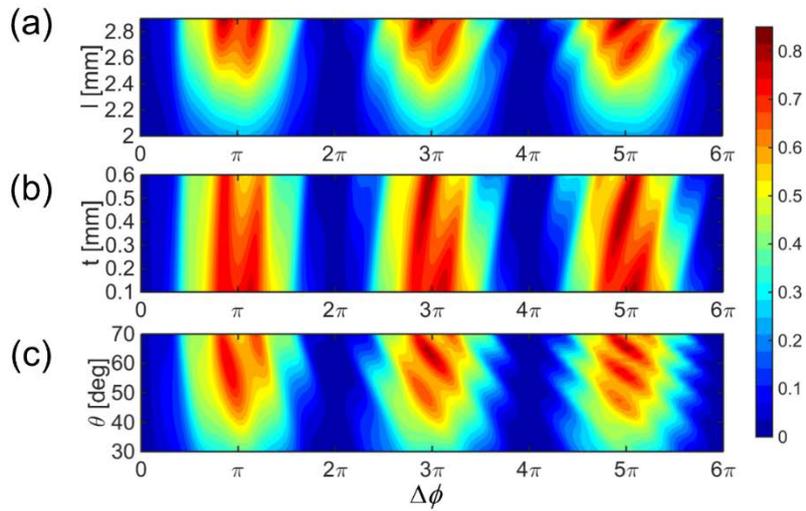

FIG. S7. Contour plots of the L-to-S mode conversion rate (a) on the $\Delta\phi - l$ plane with the fixed values of $t = 0.3$ mm and $\theta = 65°$, (b) on the $\Delta\phi - t$ plane with the fixed values of $l = 2.8$ mm and $\theta = 65°$, and (c) on the $\Delta\phi - \theta$ plane with the fixed values $l = 2.8$ mm and $t = 0.5$ mm.

# EXPERIMENTAL DETAILS

Figure S8 shows the experimental setup. The base 1 mm-thick plate is made of aluminum (Annealing 6061). The EMMs used in this study were composed of square unit cells with the lattice constant equal to 3 mm. In the frequency range of interest (around 100 kHz), therefore, the unit cell size falls into a subwavelength regime. The slits have the corner fillet of radius 200 μm due to the round laser-beam shape. The mode-coupled EMMs used for the experiments consisted of more than 1000 unit cells and had 6 to 8 unit cells along the wave propagation direction.

Both transmitting and sensing ultrasound waves were conducted by using magnetostrictive patch transducers (MPTs) [1, 3]. The detailed transducer configuration and working principle can be found in Ref. [3]. The employed transmitter is wide in the *y* direction (the direction perpendicular to the propagation direction) to generate plane and collimated ultrasonic $S_0$ waves, which is the lowest symmetric Lamb-wave mode, in a test plate. After the incident $S_0$ wave passes through the mode-coupled EMMs, it will be split into two wave modes, $S_0$ and $SH_0$ modes. The $SH_0$ mode represents the lowest shear-horizontal (SH) wave mode of the test plate. To measure the transmitted $S_0$ and $SH_0$ modes, we employed S-mode and SH-mode MPT-type receivers. The employed MPTs were so fabricated as to provide good sensitivity and selectivity for both in-plane longitudinal and shear strain excitations and measurements. In the frequency range of our interest (<150 kHz), which is far below the cutoff frequency of the higher modes (~1 MHz), both $S_0$ and $SH_0$ modes can be considered as nondispersive L and S modes, respectively, under the plane stress condition.

About the configurations of the transmitter and receivers, the size of the magnetostrictive patch of the transmitting MPT was 36 cm × 3.4 cm (width × length). If the excitation frequency is 100 kHz, for instance, the patch width becomes half the wavelength of the incident $S_0$ mode. Therefore, the selected transmitter works most efficiently around 100 kHz. However, it can be used at other frequencies even though the transmission efficiency becomes lowered. A similar type of long magnetostrictive transducers was successfully used earlier to generate $SH_0$ plane and collimated waves [4]. On the other hand, the magnetostrictive patches of the receiving MPTs are compact: 3.0 cm × 3.4 cm and 3.0 cm (or 3.4 cm) × 4.1 cm (width × length) for selectively measuring the $S_0$ and $SH_0$ modes, respectively. (The patches having the widths of 3.4 cm and 3.0 cm were used for experiments with the weakly and strongly coupled EMMs, respectively.) Because each receiver should pick up a wave signal at the center of the patch, small-sized patches are generally preferred. However, the sensitivities of too small-sized patches would be too low to be used. Also, we considered to reduce calibration errors at the resonance of the patch itself so that the patch size for the $SH_0$ mode receiver was chosen to be longer than half the wavelength of the $SH_0$ mode at frequencies around 100 kHz. In the subsequent discussions, the MPT used to transmit the S-mode wave will be called the S-mode giant MPT and those used to receive S- and SH-modes, S- and SH-mode compact MPTs, respectively.

For experiments, sinusoidal input signals were generated and amplified by a function generator (Agilent 33220A) and a power amplifier (AG1017L), respectively. Four to eleven cycles were chosen depending on the excitation frequencies. To calculate the normalized L-to-L transmission power through the EMM layer from the measured voltage of the MPT receiver, a reference signal from the S-mode giant transmitter was measured in the test plate without the EMMs. For the measurement, the S-mode compact MPT was used. On the other hand, to calculate the L-to-S conversion rate, indirect calibration of the measured $SH_0$ wave signal by the SH-mode compact MPT was necessary because there will be no SH signal sensed if there is no mode-coupled layer inserted between the S-mode transmitting giant MPT and the SH-mode compact receiving MPT (refer to Fig. S8). So, we first experimentally determined the spectrum ($\widetilde{H}_{S \to SH}^{EMM}$) in the plate with the EMM by using the SH-mode compact receiving MPT and the S-mode transmitting giant MPT. Then, we experimentally measured the spectrum ($\widetilde{H}_{S \to S}^{w/o\ EMM}$) in the test plate without any EMM installed by using the S-mode receiving giant MPT and the S-mode transmitting giant MPT. We also experimentally measured the spectrum ($\widetilde{H}_{SH \to SH}^{w/o\ EMM}$) in the test plate without any EMM installed by using the SH-mode receiving compact MPT and the SH-mode transmitting compact MPT. Finally, we determined the frequency-calibrated spectrum from $\widetilde{H}_{S \to SH}^{EMM} / \sqrt{\widetilde{H}_{SH \to SH}^{w/o\ EMM} \cdot \widetilde{H}_{S \to S}^{w/o\ EMM}}$. The least-square fitting of the frequency-calibrated spectrum was additionally conducted to compare the experimental L-to-S conversion rate with the numerical L-to-S conversion rate correctly, as shown in Fig. 4(b).

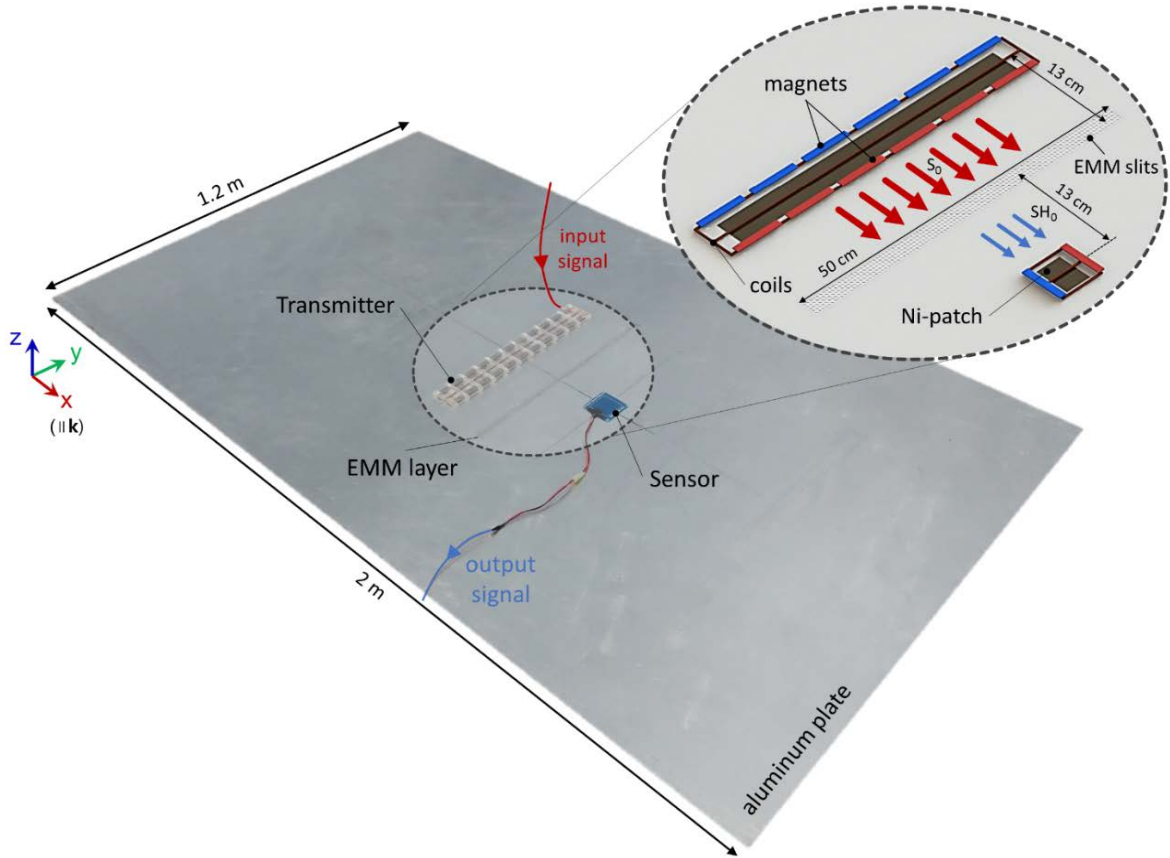

FIG. S8. Experimental setup. The mode-coupled EMM layer was fabricated in the aluminum plate. The S-mode giant MPT is used to generate an $S_0$-mode ultrasound plane collimated wave. The SH-mode compact MPT receiver is used to measure the $SH_0$ mode wave converted from the incident $S_0$-mode as it passes through the EMM layer. The inset sketches the zoomed views of the used MPTs and the mode-coupled EMM configuration.